%% file: conference_101719.tex
\documentclass[conference]{IEEEtran}
\IEEEoverridecommandlockouts
\usepackage{cite}
\usepackage{amsmath,amssymb,amsfonts}
\usepackage{algorithmic}
\usepackage{graphicx}
\usepackage{textcomp}
\usepackage{xcolor}
\usepackage{color}

\def\BibTeX{{\rm B\kern-.05em{\sc i\kern-.025em b}\kern-.08em
    T\kern-.1667em\lower.7ex\hbox{E}\kern-.125emX}}
\begin{document}

\title{Self-supervised Learning for Acoustic Few-Shot Classification\\
}

\author{\IEEEauthorblockN{Jingyong Liang}
\IEEEauthorblockA{\textit{Information Technology} \\
\textit{Monash University}\\
Melbourne, Australia \\
Jingyong.Liang@monash.edu}
\and
\IEEEauthorblockN{Bernd Meyer}
\IEEEauthorblockA{\textit{Information Technology} \\
\textit{Monash University}\\
Melbourne, Australia \\
Bernd.Meyer@monash.edu}
\and
\IEEEauthorblockN{Isaac Ning Lee}
\IEEEauthorblockA{\textit{Engineering} \\
\textit{Monash University}\\
Melbourne, Australia \\
safricanus66@gmail.com}
\and
\IEEEauthorblockN{Thanh-Toan Do}
\IEEEauthorblockA{\textit{Information Technology} \\
\textit{Monash University}\\
Melbourne, Australia \\
toan.do@monash.edu}
}

\maketitle

\input{0_abstract}


\input{1_introduction}

\begin{figure*}[t]
    \centering
    \includegraphics[scale=0.52]{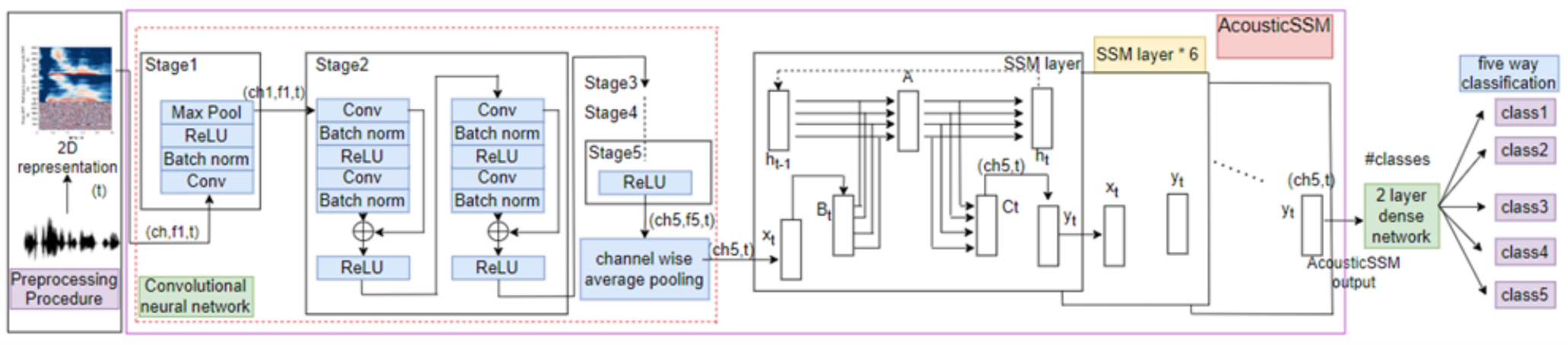}
    \caption{AcousticSSM for few-shot classification procedure.
    }
    \label{fig:systemArchitecture1}
\end{figure*}
\begin{figure}[t]
    \centering
    \includegraphics[scale=0.21]{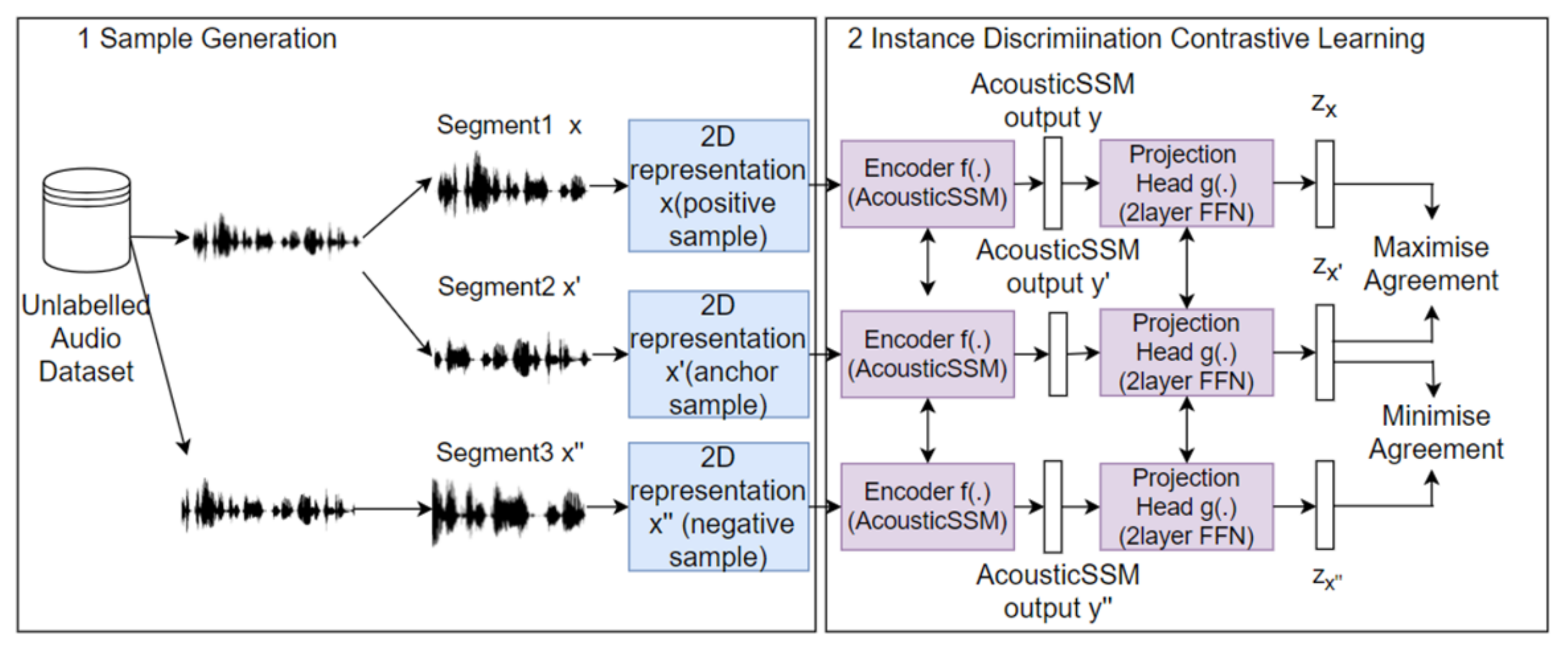}
    \caption{Contrastive training pipeline including sample generation and contrastive learning.}
    \label{fig:systemArchitecture2}
\end{figure}

\input{2_method}

\begin{table*}
    \caption{Average percentage accuracy for few-shot audio classification (AA represents average accuracy)}
    \setlength{\arrayrulewidth}{0.03mm}
    \setlength{\tabcolsep}{9.3pt}
    \renewcommand{\arraystretch}{1.2}
    
    \small
    \begin{center}
    \begin{tabular}{|c c c c c c c c c c c c|} 
    \hline
    Method & ESC50 & & & & & & & & & &   \\ [0.03ex] 
    
    \hline
           & G1 & G2 & G3 & G4 & G5 & G6 & G7 & G8 & G9 & G10 & AA\\
    \hline 
    MT-SVLR & 0.805 & 0.782 & \textbf{0.749} & \textbf{0.799} & \textbf{0.736} & 0.671 & 0.646 & 0.798 & 0.575 & 0.706 & 0.727 \\
    \hline
    Adapted CNN & 0.739 & 0.676 & 0.618 & 0.756 & 0.693 & 0.655 & 0.604 & 0.772 & 0.623 & 0.795 & 0.693 \\
    \hline
    AcousticSSM1 & \textbf{0.867}& \textbf{0.793} & 0.673 & 0.747 & 0.713 & \textbf{0.767} & \textbf{0.733} & \textbf{0.873} & \textbf{0.727} & \textbf{0.907} & \textbf{0.780} \\
    \hline
    AcousticSSM2 & 0.846 & 0.751 & 0.735 & 0.747 & 0.721 & 0.695 & 0.685 & 0.823 & 0.657 & 0.782 & 0.744 \\
    
    \hline
    \end{tabular}
    \end{center}
    \label{tab:my_label1}
\end{table*}     
\begin{table*}
    \caption{Average percentage accuracy for few-shot audio classification
    Comparison with other large-scale dataset pre-trained model (AA represents average accuracy)}
    \setlength{\arrayrulewidth}{0.030mm}
    \setlength{\tabcolsep}{6.95pt}
    \renewcommand{\arraystretch}{1.2}
    
    \small
    \begin{center}
    \begin{tabular}{|c c c c c c c c c c c c |c|} 
    \hline
    Method & ESC50 & & & & & & & & & & &Bioacoustic \\ [0.03ex] 
    
    \hline
           & G1 & G2 & G3 & G4 & G5 & G6 & G7 & G8 & G9 & G10 & AA &accuracy\\
    \hline
    Tera &  0.617 & 0.526 & 0.646 & \textbf{0.783} & 0.663 & 0.623 & 0.577 & 0.646 & 0.600 & 0.777 & 0.646 & 0.643\\ 
    \hline
    Mockingjay &  0.549 & 0.331 & 0.497 & 0.674 & 0.629 & 0.491 & 0.417 & 0.663 & 0.457 & 0.651 & 0.536 &0.586\\ 
    \hline
    Wav2vec & 0.497 & 0.291 & 0.571 & 0.669 & 0.600 & 0.474 & 0.491 & 0.640 & 0.417 & 0.577 & 0.523 & 0.721\\
    \hline 
    HuBERT & 0.554 & 0.623 & 0.709 & 0.731 & 0.686 & 0.686 & 0.657 & 0.657 & 0.606 & 0.800 & 0.671 &0.701\\
    \hline
    BYOL & 0.211 & 0.263 & 0.234 & 0.189 & 0.200 & 0.217 & 0.206 & 0.223 & 0.200 & 0.217 & 0.216 & 0.150
    
    \\
    \hline
    AcousticSSM1 & \textbf{0.867}& \textbf{0.793} & 0.673 & 0.747 & 0.713 & \textbf{0.767} & \textbf{0.733} & \textbf{0.873} & \textbf{0.727} & \textbf{0.907} & \textbf{0.780} & \textbf{0.878}\\
    \hline
    AcousticSSM2 & 0.846 & 0.751 & \textbf{0.735} & 0.747 & \textbf{0.721} & 0.695 & 0.685 & 0.823 & 0.657 & 0.782 & 0.744 &0.864\\
    
    \hline
    \end{tabular}
    \end{center}
    \label{tab:my_label2}
\end{table*}

\input{3_experiment}
\input{4_conclusion}

\end{document}

%% file: 0_abstract.tex
\begin{abstract}
Labelled data are limited and self-supervised learning is one of the most important approaches for reducing labelling requirements. While it has been extensively explored in the image domain, it has so far not received the same amount of attention in the acoustic domain. Yet, reducing labelling is a key requirement for many acoustic applications. Specifically in bioacoustic, there are rarely sufficient labels for fully supervised learning available. This has led to the widespread use of acoustic recognisers that have been pre-trained on unrelated data for bioacoustic tasks. We posit that training on the actual task data and combining self-supervised pre-training with few-shot classification is a superior approach that has the ability to deliver high accuracy even when only a few labels are available. To this end, we introduce and evaluate a new architecture that combines CNN-based preprocessing with feature extraction based on state space models (SSMs). This combination is motivated by the fact that CNN-based networks alone struggle to capture temporal information effectively, which is crucial for classifying acoustic signals. SSMs, specifically S4 and Mamba, on the other hand, have been shown to have an excellent ability to capture long-range dependencies in sequence data. We pre-train this architecture using contrastive learning on the actual task data and subsequent fine-tuning with an extremely small amount of labelled data. We evaluate the performance of this proposed architecture for ($n$-shot, $n$-way) classification on standard benchmarks as well as real-world data. Our evaluation shows that it outperforms state-of-the-art architectures on the few-shot classification problem. 
\end{abstract}

%% file: 1_introduction.tex
\section{Introduction}
Reducing labelling requirements is central to many application areas since obtaining labelled training data usually requires extensive human labour and is thus costly and error-prone. This is specifically true in the application domain we are targeting, bioacoustics, where the cost of extensive labelling is often prohibitive\cite{b1}.

One of the most promising approaches for reducing labelling requirements is the use of self-supervised learning\cite{b2,b3}. Self-supervision can be used in an initial pre-training phase and the networks obtained can subsequently be fine-tuned with a very small amount of labelled data for specific downstream tasks. We are specifically concerned with classification, where the above approach is the basis of $n$-way, $n$-shot classification in which just $n$ labelled samples per class are used to train a classifier for $n$ way. 

Such $n$-way, $n$-shot classification is specifically relevant in bioacoustic. Here, a typical task is the recognition of threatened species\cite{b4}. Since sufficient problem-specific labelled train data is rarely available, particularly for cryptic species, the use of recognisers that have been pre-trained on unrelated data is general practice. For example, mammals are commonly classified with recognisers that have been pre-trained on bird data or even generalised audio data, such as AudioSet\cite{b5}.

It stands to reason that pretraining feature embedders on the actual problem data should lead to superior performance, since it specialises the embedder for the specific setting. However, without extensive labelling efforts, this has to be achieved with self-supervised (or unsupervised) methods. Unfortunately, self-supervised learning has not yet been as widely explored in acoustic processing as in image processing\cite{b6} and architectures for self-supervised image processing cannot necessarily directly be transferred to the acoustic domain. This is because acoustic data is fundamentally different from image data in that it constitutes sequence data with a temporal dimension. Thus, a self-supervised architecture suitable for audio processing is best based on models for sequence data. In the present paper, we propose such an architecture.

Our architecture is based on a combination of convolutional neural network (CNN) blocks for feature pre-processing with a Structured State Space Sequence model (S4). The use of S4 is motivated by the fact that S4 architectures, including Mamba, are specifically designed to model long sequences and have been proven to have great potential for modelling long-range dependencies in sequence data. SSMs reach equivalent performance to other sequence models, specifically Transformers, while significantly reducing the computational effort\cite{b7,b8,b9,b10}. CNNs, on the other hand, are the most commonly used approach for supervised learning of audio data\cite{b11} and are effective at feature processing. We thus use a CNN structure based on ResNet\cite{b12} for initial feature processing, leaving the time-dimension untouched, and subsequently process the time series of preprocessed features with an S4 architecture. We specifically use S4D\cite{b13,b14}, an improved version of S4 that reduces the computational requirements. 

We evaluate the use of our architecture for five-way five-shot classification on standard benchmark data (ESC 50) as well as on real-world data recorded for 10 different frog species recorded in the field in Queensland, Australia\cite{b15}. We pre-train the embedder on the actual problem data and subsequently fine-tune the downstream classifier with labelled samples from the same data set. This truthfully reflects the real application scenario, where unlabelled data from the specific application scenario is generally available in copious amounts while labelling is very expensive so that only a few labelled samples are available. It is important to note that this is subtly different from other versions of the $n$-way, $n$-shot few-shot classification, where data used in the pre-training are different from those used for the fine-tuning \cite{b16}. The latter tests the generalisation ability of the network to new unseen classes, whereas we are interested in achieving increased performance by specialising an embedder for a specific application scenario using only a very limited labelling budget. 

We evaluate the performance of our self-supervised model to the state-of-the-art models of two different approaches: we compare (a) to other self-supervised classification models that are trained in the exact same way as our model (self-supervised pre-training on unlabelled problem data, subsequent fine-tuning with labelled samples) and (b) to feature embedders that have been pre-trained on large amounts of unrelated data and then fine-tuned with problem-specific data. We use the same labelling budget for both approaches. In both cases and for both types of benchmark datasets we find that our architecture outperforms the state of the art.

The core contributions of this paper as thus twofold: Firstly, we introduce a new architecture for self-supervised training on audio data that exceeds state-of-the-art performance. Secondly, we utilise this architecture for a methodology that allows us to specialise robust feature embedders (that can be used for various downstream tasks) for specific application scenarios, using only very few labelled samples.

%% file: 2_method.tex
\section{Acoustic SSM for few-shot classification}
Fig.~\ref{fig:systemArchitecture1} illustrates the proposed architecture. We first perform audio preprocessing for the raw input waveform and then transform the waveform into stacked spectrograms, consisting of a Mel spectrogram and the Short-Time Fourier Transform (STFT) magnitude and phase angle. The output shape after preprocessing is $(ch, f, t)$ where $ch=3$; $f$ and $t$ indicate the frequency and the time dimensions, respectively. 
After that, the spectrograms are processed by our proposed feature extractor -- the AcousticSSM, which consists of CNN and SSM components. The spectrograms are first processed by a residual CNN, capturing dependencies in the frequency dimension while keeping the time dimension untouched. 
Next, the extracted features are processed by SSM blocks, which extract features by learning long-range sequence  dependencies in the temporal dimension without changing the frequency channels. The SSM output is a latent space used for contrastive learning and, subsequently, classification. This structure is illustrated in Fig.~\ref{fig:systemArchitecture1}.
\subsection{Audio feature embedder - AcousticSSM}
We design our convolution block based on the residual block proposed in ResNet\cite{b12}.  We design this block to extract only features within the frequency domain that are local in the time domain and repeat the process along the time dimension. The first convolution stage consists of a convolutional layer containing $ch1=64$ filters of spatial size of $7\times 1$, followed by a batch normalisation (BN) layer, a ReLU, and a $3\times1$ max pooling layer. Each of the next four stages identically repeats a structure of two residual blocks (Fig.~\ref{fig:systemArchitecture1}). 
Each residual block consists of two convolutional layers. The spatial size of the filter in every convolutional layer is $3\times1$. The number of filters in each convolutional layer of stage numbers $i=2,\ldots,5$ is $ch_i=2^{4+i}$.
Channel-wise $f\times 1$ average pooling is then applied to the frequency dimension, transforming the data dimensions from $(ch5,f5,t)$ to $(ch5,t)$.

State space models (SSM) stem from classical continuous system dynamics and map a one-dimensional input function $x(t) \in {\cal R}$, through intermediate hidden states $h(t) \in {{\cal R}^N}$ to outputs $y(t) \in R$\cite{b7,b8}. Effectively, an SSM models a linear ODE with learnable parameters A, B, and C as shown below:
\begin{equation}
    h'(t) = Ah(t) + Bx(t)
\end{equation}
\begin{equation}
    y(t) = Ch(t).
\end{equation}
After extracting frequency features with the CNN, we use six sequential SSM layers to extract long-range dependencies between features across the temporal dimension. In each SSM layer, multiple SSMs work independently on the channels in parallel.
Every SSM layer maintains its input shape.

\subsection{Contrastive learning method}
Motivated by SimCLR\cite{b17} and COLA\cite{b18}, we learn a robust representation of unlabelled audio signals by training our proposed AcousticSSM with a contrastive loss function. We investigated two different contrastive learning methods in our experiments. The first method (AcousticSSM1) assigns high similarity to audio segments extracted from the same audio recording and low similarity to audio segments extracted from different audio clips. The loss function maximises the agreement between an anchor segment and a related positive segment from the same audio clip while minimising the agreement between this anchor segment and negative segments from unrelated clips. Instead of keeping a memory bank of negatives after picking one anchor segment, we use a bullet strategy: positive segments are defined as negative segments for all other anchors in one batch.

For the second contrastive learning method (AcousticSSM2), we apply augmentations to generate related samples. Motivated by CLAR\cite{b19}, we use augmentation that combines pitch shift, fade in/out, time masking, and time shift. In this case, the anchor example and its corresponding related positive sample come from the same extracted audio segment with different augmentation parameters, whereas the unrelated negative samples are augmented audio segments from different clips. 

For both contrastive learning methods, we use the same contrastive loss function calculated with the pipeline shown in Fig.~\ref{fig:systemArchitecture2}. First, an encoder $f$ (here, the proposed AcousticSSM) is applied to map the preprocessed input into a latent representation $y = f(x)$. The latent representation $y$ is used for the downstream few-shot classification tasks after pre-training. 
After that, a projection head $g$ maps $y$ into a latent feature $z$, i.e., $z=g(y)$, where the projection head is a two-layer MLP layer. Motivated by COLA\cite{b18}, we use bilinear learnable parameters $W$ to calculate the similarities between two audio samples $(x,x')$ as follows:
\begin{equation}
    s(x,x') = {g(f(x))^T} W g(f(x')).
\end{equation}
After calculating the similarities, we apply a multi-class cross-entropy loss:
\begin{equation}
    L = -log \frac{exp(s(x,x^+))}{\sum\limits_{x_i \in \mathcal{N}(x) \cup \{x^+\}}{exp(s(x,x_i))}},
\end{equation}
 where ${x^+}$ represents the positive sample associated with anchor $x$ and $\mathcal{N}(x)$ represents the set of negative samples corresponding to $x$.

\subsection{Few-shot downstream classification task}
To address our few-shot classification task, we transfer the pre-trained AcousticSSM model to the downstream classifier. The output of the AcousticSSM encoder is the latent representation $y=f(x)$, as mentioned in the previous section. This is fed into a two-layer dense network to obtain a classifier for the few-shot audio classification task using cross-entropy loss function. The pipeline is shown in Fig.~\ref{fig:systemArchitecture1}. In our experiment, we focus on the five-way five-shot task.

%% file: 3_experiment.tex
\section{Experimental Evaluation}

We evaluate our method on the well-known Environmental Sound Classification benchmark ESC50\cite{b20} as well as on a real-world bioacoustic dataset. To the best of our knowledge, no existing benchmark for the ESC50\cite{b20} five-way five-shot classification tasks is available. 
ESC50 consists of 2,000 5-second samples of environmental recordings equally distributed across 50 classes (40 clips per class). Our bioacoustic dataset contains calls of 5 frog species recorded at 4-Mile-Creek, Townsville, Queensland, Australia in 2020 \cite{b15} ({\it L rubella}, {\it L rothii}, {\it L pornatum}, {\it C novae}, {\it L caerulea}).\footnote{Thanks to Lin Schwartzkopf and 
Slade Allen-Ankins (James Cook University, Townsville) for making this dataset available.} We use 50 unlabelled samples of each species and 5 labelled samples for fine-tuning.
To fit our 5-way 5-shot 
 problem, we group ESC 50 into 10 groups $G_1, \ldots, G_{10}$ with each group containing 5 classes and group divisions according to coarse semantic categories (animal vocalisations, human vocalisations, natural environment sounds, interior built environment, urban exterior sounds, ...).

All recordings were resampled at 20kHz with length 30225 ($\sim$1.5s) and all audio samples were pre-processed using cropping, padding, and augmentation according to\cite{b21}.

In each experiment, one group $G_i$ is chosen. For each of the 5 ways in $G_i$ 5 labelled samples are reserved for fine-tuning and all other samples (but without labels) are used for pre-training. We use ADAM for pre-training and SGD for fine-tuning. For pre-training, we apply a learning rate of 0.0001 for 500 epochs. For fine-tuning, we apply a learning rate of 0.006 for 50 epochs, where the best parameter was optimised using a parameter sweep.
\footnote{The learning rate for fine-tuning was optimised using a parameter sweep between 0.005 and 0.01.}

Table~\ref{tab:my_label1} compares our approach to MT-SVLR, a current self-supervised SOTA model\cite{b22}, trained in the exact same way.
We also compare the performance of our 
approach to the alternative of fine-tuning a network that received self-supervised pre-training on another (unrelated) dataset not directly taken from the targeted application. These models are pre-trained on the large-scale sample set LibriSpeech~\cite{b23}, which is provided by the SUPERB benchmark\cite{b24}. Table~\ref{tab:my_label2} compares our method to Tera\cite{b25}, Mockingjay\cite{b2}, Wav2vec\cite{b26}, HuBERT\cite{b27} and BYOL\cite{b28}.
As using a pre-trained classifier is common practice in bio-acoustics, we also report the results for our bioacoustic real-world data in this context. 

\subsection{Results}
\label{sec:typestyle}
From Table.~\ref{tab:my_label1} it is evident that the AcousticSSM outperforms MT-SLVR, representing current SOTA, on average and for the majority of test groups. For the average this is true regardless of the version of contrastive learning. Overall AcousticSSM1 shows superior performance. 

To confirm our hypothesis that the sequence processing abilities of the SSM component are crucial, we perform an ablation study in which we only use the CNN component of the combined model (``Adapted CNN'' in Table~\ref{tab:my_label1}).
The AcousticSSM significantly outperforms the adapted CNN model, confirming our hypothesis.  
This highlights the importance of learning temporal long-range features by utilizing an SSM-based network.

Both versions of AcousticSSM also clearly outperform the models that transferred from pre-training on another large-scale dataset (Table~\ref{tab:my_label2}). The next-best performing model is HuBERT with a performance gap of $>$10\% to AcousticSSM1. The test on our bioacoustic dataset shows the same trends, with the AcousticSSM significantly outperforming all other methods. Interestingly, while all other methods exhibit comparable average performance for ESC50 and for the bioacoustic data, wav2vec shows a much-improved performance for the real-world data. Yet, it still remains far behind the results of the AcousticSSM with a performance gap of $>$15\% on average and $>$25\% for the real-world data.

%% file: 4_conclusion.tex
\section{Conclusion}
This paper introduced and evaluated a new self-supervised approach for few shot acoustic classification tasks. Our AcousticSSM architecture serves as a feature extractor for acoustic data and is based on a 
combination of a feature pre-processing CNN with a state space model-SSM, which is utilized to help extract deep and robust features of the sequence input signal by learning a long range dependencies in temporal dimension.

The experimental evaluation shows that our AcousticSSM  learns high performing acoustic  feature extractors that enable higher than SOTA accuracy in the downstream classification 
task. It also 
shows that our self-supervised method (pre-training feature embedders on the actual problem data and then fine-tuning on a very small amount of labelled data) facilitates superior performance.

This approach truthfully reflects the requirements of applications in which 
unlabelled data from the specific scenario is available in large amounts while labelling is very expensive. This indicates that our method could be transferred to other  application cases with similar characteristics.

Going forward, AcousticSSM could be applied as a feature extractor in the acoustic domain and even the video domain due to its competitive ability to capture long range sequence dependencies. We expect that such self supervised feature extractors can be useful to improve a broad range of downstream tasks, including event detection.